# Singlet Stripe Phases in the planar $t$-$J$ Model

Hirokazu Tsunetsugu[1,2], Matthias Troyer[1,2,3], and T. M. Rice[1]

[1] *Theoretische Physik, Eidgenössische Technische Hochschule, 8093 Zürich, Switzerland*
[2] *Interdisziplinäres Projektzentrum für Supercomputing, ETH-Zentrum, 8092 Zürich, Switzerland*
[3] *Centro Svizzero di Calcolo Scientifico, 6924 Manno, Switzerland*

(July 1994)

The energies of singlet stripe phases in which a plane is broken up into spin liquid ladders by lines of holes, is examined. If the holes were static then patterns containing spin liquids with a finite spin gap are favored. The case of dynamic holes is treated by assembling $t$-$J$ ladders oriented perpendicular to the stripes. For a wide region around $J/t \approx 1$ the hole-hole correlations in a single ladder are found to be predominantly charge density wave type but an attraction between hole pairs on adjacent ladders leads to a stripe phase. A quantum mechanical melting of the hole lines at smaller $J/t$ values leads to a Bose condensate of hole pairs, i.e. a superconducting phase.

Although there are many aspects of the planar $t$-$J$ model that are by now well established, such as the existence of a region of $d_{x^2-y^2}$-pairing at low doping and moderate $J/t$ ratios[1] and the occurrence of some form of phase separation (PS) at larger values,[2] $J/t \gtrsim 1$, there are also many unresolved issues. Most importantly a clear physical picture is lacking which would interrelate and give insight into features such as the very rapid suppression of long range antiferromagnetic (AF) order with doping, the absence of long range incommensurate antiferromagnetism (ICAF), the anomalous boundary of the PS regime extending down to values $J/t \sim 1$ (e.g. see Prelovšek and Zotos[3]), and of course the presence of $d_{x^2-y^2}$-pairing. Similarly the phenomenological nearly-AF liquid models studied by Moriya, Takahashi and Ueda,[4] and by Monthoux, Pines and coworkers[5] can explain the high $T_c$ values but only if there is an extended critical regime with very large enhancements of the staggered susceptibility which for some unspecified reasons do not crossover to long range magnetic order (for a review of these issues see Ref.[6]). In this communication we shall address these points by studying the dimensional crossover from ladders to planes in the $t$-$J$ model.

We start with $t$-$J$ ladders rather than chains. Both are one dimensional but nonetheless behave very differently — a point which alone emphasizes the subtlety of these issues. The properties of $t$-$J$ chains are well known. They form Luttinger liquids upon hole doping with correlation exponents $K_\rho \approx \frac{1}{2}$ for $J/t \lesssim 1$.[7] The dominant or longest-range correlations are incommensurate magnetic correlations and there is no sign of pairing. Coupling such chains together to make a crossover to a planar model would seem to lead to ICAF phases contrary to both experiments and to studies of the planar $t$-$J$ model. Note that a sequence commensurate-AF → ICAF → paramagnetism appears elsewhere, most notably as the electron/atom ratio is changed in the Cr alloys.[8] So why is the physics different here?

Recently it has been discovered that the properties of $t$-$J$ ladder systems can be very different and depend sensitively on the width of the ladders.[9] Ladders made from an even number of chains (or legs) are in the Luther-Emery class with a finite spin gap and exponential decay of the spin-spin correlations. This points to unexpected subtleties in the behavior of the magnetic order in this extreme quantum system. First we examine the energies involved in breaking up the infinite plane into ladder stripes. A key question here is whether the formation of spin liquids which lead to a spin gap is favored. We will examine this question first for static holes for configurations with a fixed density of broken bonds. In the $t$-$J$ model the holes of course are not static but highly dynamic depending on $J/t$. At larger values of $J/t$ the holes are less mobile and here we look at the energetics of stripe phases but with singlet spin liquids formed between the hole stripes. Such singlet stripe (SST) phases can be viewed as an assembly of ladders perpendicular to the stripes. If the hole pairs on adjacent ladders are attracted to each other, then depending on the correlation exponent, $K_\rho$, of the density-density correlation function on individual ladders, a SST phase will be stabilized. The alternative phase for hole pairs moving in the spin liquid background is a Bose condensed state, i.e. the singlet superconducting (SSC) phase. The transition between SST and SSC order we view as a quantum melting transition which can occur as $J/t$ changes.

The first question is the relative stability of the spin liquid (short range RVB) and the spinon phases. Analogously to interacting fermion systems we expect the opening of the spin gap in the short range RVB phase to be accompanied by a gain in energy. In the present context we compare ground state energies of two static hole configurations at hole doping $\delta = 0.25$. One is a charge density wave (CDW) with lines of holes separating 3-chain spin ladders and the second has alternating 2-chain and 4-chain spin ladders (see Fig. 1). Both configurations have an equal number of broken magnetic bonds (3 per hole) so we can directly compare them. The energies per site for 2-, 3-, and 4-chain ladders quoted in Table 1 were obtained by extrapolating the results of Lanczos diagonalizations to infinite ladders. The direct comparison of the energies of an RVB spin liquid with 2- and 4-chain ladders and a spinon fluid with 3-chain ladders shows a small energy gain for the former of $\approx 6 \times 10^{-3} J$/site or $\approx 1\%$ of the magnetic energy.

So far we have set $t = 0$. Turning on $t$ will lead to fluctuations in the hole lines. The increasing width of the lines can be, partly at least, accounted for in a variational Monte Carlo (VMC) treatment of a projected wave function incorporating both CDW and $d$-wave RVB correlations. The spin gap in this system would arise when the pairing amplitude has different signs on different Fermi surface sheets but no nodes at the Fermi surface. Viewed in this way the stability of the SST phase comes from their absences of nodes in the gap function.

Here however we use a different approach to examine the stability of SST phases. We break up the system into 2-chain ladders running perpendicular to the stripes and then examine the crossover from ladders to a 2D plane as we turn on the interladder terms. First we review the properties of single (isotropic) ladders in the $t$-$J$ model which has the standard form,

$$H = -t \sum_{\langle i,j \rangle, \sigma} \left( \tilde{c}_{i\sigma}^\dagger \tilde{c}_{j\sigma} + \text{H.c.} \right) + J \sum_{\langle i,j \rangle} \left( \mathbf{S}_i \cdot \mathbf{S}_j - \tfrac{1}{4} n_i n_j \right), \quad (1)$$

with $\tilde{c}_{i\sigma} = c_{i\sigma}(1 - n_{i,-\sigma})$ etc. The undoped ladder is a spin liquid with a spin gap $\simeq J/2$.[10] Upon doping the results of Lanczos diagonalization studies show that the spin gap remains but changes discontinuously due to the introduction of new states associated with breaking up hole pairs into two separated charged spin-$\tfrac{1}{2}$ quasiparticles.[9] Doped ladders fall into the Luther-Emery class of 1D Fermion models. The density-density and pair-pair correlation functions fall off with power law and the exponents are determined by the parameter $K_\rho$ whose value is given by $K_\rho = \pi \rho (\kappa D/2)^{1/2}$ where $\rho$ is the electron density, $\kappa$ is the compressibility and $D$ is the Drude weight.[11] The Drude weight is calculated by evaluating the dependence on a phase shift of the boundary conditions (BC) of the ground state energy for a $2 \times L$ ladder with 2 holes. In calculating the compressibility $\kappa$, care must be taken since the energy of a finite ladder with periodic BC's depends sensitively on whether an AF spin arrangement along the chains is frustrated or not, e.g. 2 holes in $2 \times 8$ is frustrated while 2 holes in $2 \times 7$ is not. We calculate $\kappa$ from the ground state energies of unfrustrated ladders with periodic BC's (0 hole in $2 \times 8$, 2 holes in $2 \times 7$, and 4 holes in $2 \times 8$) and determine the PS boundary by the condition $\kappa \to \infty$, as shown in Fig. 2(a). Further points on PS boundary line were determined with open BC's to avoid frustration effects. Note the critical value for $(J/t)_{PS}(\approx 2.2)$ is considerably larger than that reported for the 2D case $(J/t)_{PS} \simeq 1$.[2] The corresponding values of $K_\rho$ are shown in Fig. 2(b). The value $K_\rho = 1$ separates a predominantly CDW region $(K_\rho < 1)$ from the region $(K_\rho > 1)$ where SSC correlations dominate. We note that $K_\rho \approx 0.5$ for $J/t \sim 0.3$ which agrees with the results of Noack et al[12] who used a density matrix renormalization group method to directly calculate the correlation functions of a Hubbard ladder.

Next we consider the effect of turning on the interaction between the ladders so as to crossover to the 2D or planar limit. Efetov and Larkin discussed within a mean field (MF) treatment the competition between SSC and CDW phases when a weak coupling between 1D Luther-Emery systems is switched on.[13] The corresponding MF transition temperatures $T_c^\alpha$ ($\alpha$: SSC or SST) is determined by the condition

$$1 = W^\alpha \int dR\, d\tau\, G^\alpha(R, \tau). \quad (2)$$

For $\alpha$=SSC, $W^\alpha$ is twice the pair hopping amplitude, $2t_{\text{eff}}$, and $G^\alpha$ the pairing correlation function. A recent study of a 4-chain ladder by Poilblanc et al[14] with 2 holes shows an energy splitting in the singlet sector between states with even and odd parities w.r.t. the mirror symmetry that exchanges 2-chain ladders and we set $W^{\text{SSC}}$ equal to this splitting (see Fig. 3). In Fig. 4 we show $T_c^{\text{SSC}}$ from Eq. 2 using $K_\rho$ and $v_c$ values from Fig. 2(b). The second possibility, namely CDW ordering in the case that hole pairs on adjacent ladders attract is similar to the stripe phases discussed previously in Hartree-Fock calculations of the Hubbard model.[15] An important difference is that here the spins prefer to form a spin liquid as discussed earlier giving SST rather than ICAF magnetic order. In this case $W^{\text{SST}}$ is determined by the interaction between hole pairs on adjacent ladders. This has an attractive magnetic contribution consisting of one less broken bond, $E_J^{\text{bond}} = -0.6055 J$. However separated hole pairs gain energy by a polarization process through virtual hopping which we estimate roughly to be, $P = \Delta E^{2L} - \Delta E^{4L} - E_J^{\text{bond}}$, where $\Delta E^{nL}$ is energy difference between 0- and 2-hole ground states for $n$-chain ladders (for $n = 4$ we take the average of even and odd parity). The magnetic and polarization contributions are plotted in Fig. 3 and show a change from repulsion to attraction when $J/t \gtrsim 1$. This value is an underestimate we believe since we have omitted the polarization corrections to the attractive term. Equation 2 with $G^{\text{SST}}$ as the Fourier component of the density-density correlation function gives a MF $T_c^{\text{SST}}$ shown in Fig. 4. We see at once that it dominates for $J/t \gtrsim 1$, leading us to conclude that there can be a wide parameter region of stability of the SST phase up to phase separation at $J/t \gtrsim 2.2$. In the region $J/t \lesssim 1$ where interladder interactions are repulsive one should also consider the possibility of a crystalline ordering of hole pairs, but we believe such a crystalline order would suppressed by quantum fluctions.

Following this line of reasoning leads us to a phase diagram with an intermediate parameter range $1 \lesssim J/t \lesssim$

2.2 in which SST phases are stable. This range is bounded by true PS at larger $J/t$ and SSC with $d_{x^2-y^2}$-pairing at smaller values. This form is close to that proposed by Prelovšek and Zotos[3] on the basis of the hole-hole correlations in small clusters. The key difference is that here we would have an inhomogeneous phase which is a pure CDW and the spins in the intervening regions between hole lines form spin liquids. This is the essence of the SST phase. The existence of SST phases gives an explanation for the issues raised earlier namely the anomalous PS line, the absence of long range ICAF order, and the existence of $d_{x^2-y^2}$-pairing as a precursor to the SST phase at large $J/t$ values.

Another point we would like to make concerns the anisotropic form of the SST phase. An isotropic arrangement of the hole lines is of course possible and may well be preferable. This would correspond to hole lines $\parallel x$ and $\parallel y$ axes or in other words a singlet square (SSQ) arrangement. The relative stability of SST and SSQ phases can be addressed using VMC calculations and will require further study. It is clear that small energy differences can determine the relative stability of different SST and SSQ configurations and a complex phase diagram with different patterns depending on the doping $\delta$ and the residual interactions can result. Also the energy differences to the ICAF stripe phases[15] will be very small too. The point that we want to make is that when we look at inhomogeneous phases there is good reason to believe that the SST or SSQ phases have lower energy because as shown earlier in Table 1, if one breaks up a plane the magnetic energy favors the formation of spin liquid phases. Another way of viewing the competition between SST (or SSQ) and ICAF phases is that the spin gap in the magnon spectrum should remain finite even when one includes the induced interaction across the hole lines. If the magnon energy were to go negative at some $\mathbf{q}$ vector then an ICAF phase would result. The stability of SST (or SSQ) phases require that the magnon energy remains finite leading to a spin gap and finite range spin correlations.

In conclusion we have examined the crossover between coupled $t$-$J$ ladders and the 2D limit of a $t$-$J$ plane as the interladder coupling is increased. Isolated ladders in the parameter range $J/t \approx 1$ have predominantly repulsive correlations between hole pairs leading to longer range CDW correlations but at the same intermediate parameter range the force between hole pairs on adjacent ladders is attractive. This combination of attraction and repulsion we propose stabilizes the SST (or more generally SSQ) phases. In the Fermion MF treatment these SST (or SSQ) phases have combined $d$-wave RVB pairing and CDW ordering. In that sense they are a form of supersolid. As the ratio $J/t$ is lowered the attraction between hole pairs on adjacent ladders is reduced, the quantum fluctuations of the hole lines are enhanced and at end we have a quantum melting of the CDW ordering but not of the $d$-wave RVB pairing leading to a $d$-wave superconducting state.[16] Finally we speculate that the observation of a spin gap in underdoped cuprates could result from residual CDW fluctuations which may also be precursors to true SST (or SSQ) phases.

The authors thank Didier Poilblanc for providing his data of the 4×8-site system. They also thank Peter Prelovšek and Hartmut Monien for valuable discussions. The work was supported by the Swiss National Science Foundation Grant No. NFP-304030-032833 and by an internal grant of ETH-Zürich.

TABLE I. Ground state energies per site for the Heisenberg ladders with 3 and 4 legs. Shown are the energies for a finite ladder with length $L=4$, 6 and 8 and the extrapolated value for an infinite ladder. The data (a) is taken from Ref. 14. The extrapolated value for the simple ladder (b), taken from Barnes et al. in Ref. 10 is included for reference.

| $L$ | $E_g(3\times L)/J$ | $E_g(4\times L)/J$ | $E_g(2\times L)/J$ |
|---|---|---|---|
| 4 | $-0.62751$ | $-0.64152$ | — |
| 6 | $-0.61036$ | $-0.62606$ | — |
| 8 | $-0.60568$ | $-0.62210$[a] | — |
| $\infty$ | $-0.601$ | $-0.622$ | $-0.578$[b] |

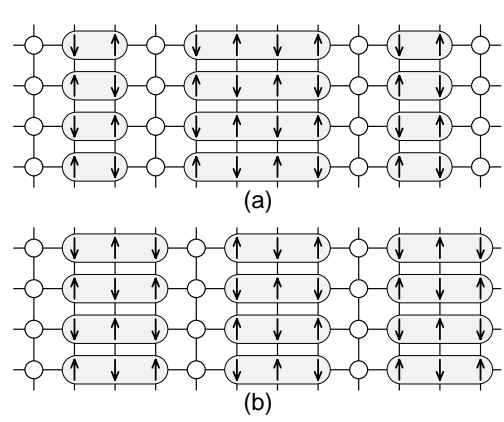

FIG. 1. Stripe phases. (a) Mixture of 2-chain ladders and 4-chain ladders. (b) 3-chain ladders.

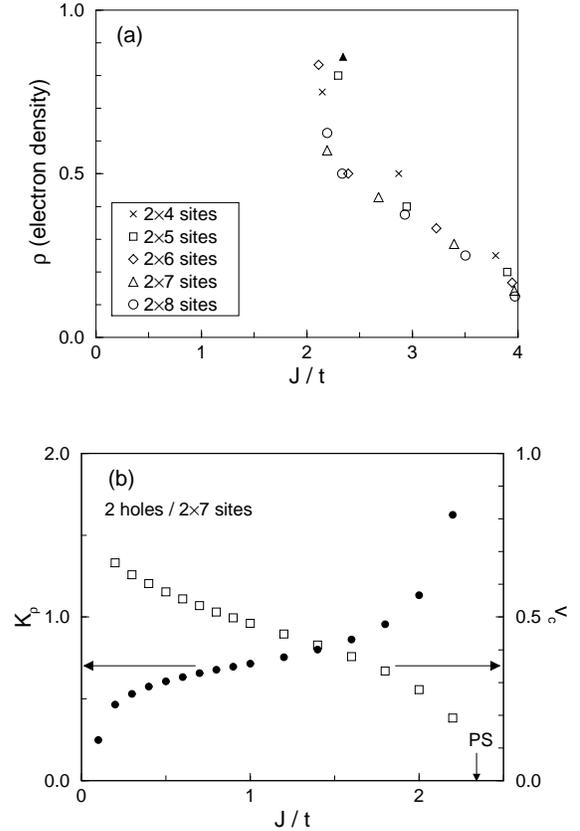

FIG. 2. (a) Phase separation boundary of $t$-$J$ ladders. Open BC's are used. ▲ is determined by the data for 2 holes/$2\times 7$, $4/2\times 8$, and $0/2\times 8$ with periodic BC's as discussed in the text. (b) Correlation exponent parameter $K_\rho$ and charge velocity $v_c$. The compressibility used here is calculated from the data for unfrustrated systems with periodic BC's.

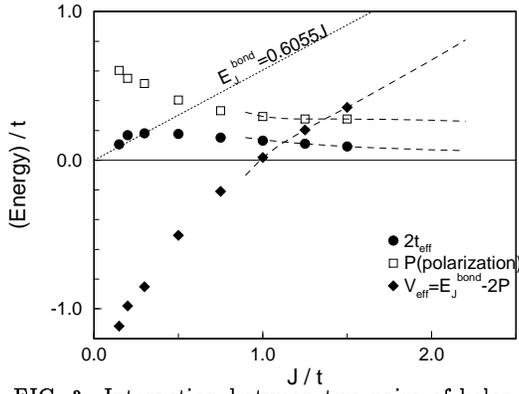

FIG. 3. Interaction between two pairs of holes determined by using data in Ref. 14. $V_{\text{eff}}$ is the effective attraction between adjacent ladders. The dashed lines are extrapolated values.

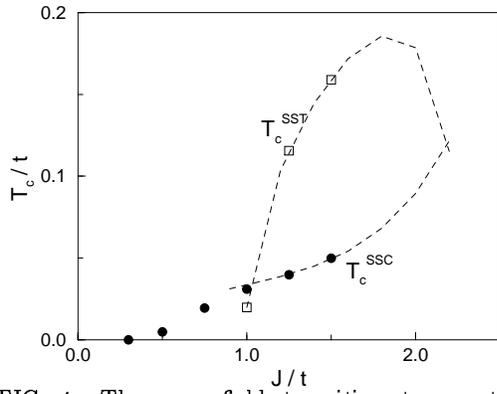

FIG. 4. The mean-field transition temperatures. The dashed lines are extrapolated values.